\documentclass[twocolumn,showpacs,amsmath,amssymb,prl]{revtex4}
\usepackage[dvips]{graphicx}

\newcommand{\norm}[1]{\left\Vert#1\right\Vert}

\newcommand{\bkt}[1]{\left\langle#1\right\rangle}
\newcommand{\up}{\uparrow}
\newcommand{\dn}{\downarrow}
\newcommand{\bs}{\backslash}
\newcommand{\sgn}{{\mathbf{sgn}}}
\newcommand{\La}{\Lambda}
\newcommand{\Ne}{{N_\mathrm{e}}}
\newcommand{\PhiGNe}{\Phi_{\mathrm{G},N_\mathrm{e}}}

\newcommand{\qed}{\rule{6pt}{6pt}}
\newcommand{\calK}{\mathcal{K}}
\newcommand{\rmi}{\mathrm{i}}
\newcommand{\rme}{\mathrm{e}}
\newcommand{\order}{\Delta_{\Lambda}}
\newcommand{\orderd}{\Delta_{\Lambda}^\dagger}
\newcommand{\cis}{c_{i,\sigma}}
\newcommand{\cisd}{c_{i,\sigma}^\dagger}
\newcommand{\nis}{n_{i,\sigma}}

\newcommand{\axs}{a_{x,\sigma}}
\newcommand{\axsd}{a_{x,\sigma}^\dagger}
\newcommand{\taxs}{\tilde{a}_{x,\sigma}}

\newcommand{\bus}{b_{u,\sigma}}
\newcommand{\busd}{b_{u,\sigma}^\dagger}
\newcommand{\Hbhop}{H_{\mathrm{hop}}^b}
\newcommand{\Hahop}{H_{\mathrm{hop}}^a}
\newcommand{\Hint}{H_\mathrm{int}}
\newcommand{\Hintp}{H_\mathrm{int}^\prime}
\begin{document}

\title{ 
 Exact Electron-Pairing Ground States of 
 Tight-Binding Models with Local Attractive Interactions  
}
\author{Akinori Tanaka}
\affiliation{%
Department of Applied Quantum Physics, Kyushu University,
Fukuoka 812-8581, Japan}
\date{\today}
\begin{abstract}
We present a class of exactly solvable models of correlated electrons.
The models are defined in any dimension $d$ and
consist of electron-hopping terms and 
local attractive interactions between electrons.
For each even number of electrons less than or equal to $1/(d+1)$-filling,  
we find the exact ground state 
in which all electrons form pairs
of a certain type, and thus the models exhibit an electron-pair condensation.
\end{abstract}
\maketitle
The origin of high-temperature superconductivity
has attracted much interest and
is still controversial.
On the other hand, the properties of
usual (low-temperature) superconductors are
well explained by the celebrated 
Bardeen-Cooper-Schrieffer (BCS) theory~\cite{BCS57}.
On the basis of the BCS theory, 
one treats
electron systems with effective 
electron-electron attraction arising
{from} electron-phonon couplings and  
discusses a phase transition associated with 
an electron-pair condensation  
by using mean-field type approximations.
However,   
whether attractive interactions between electrons 
really induce 
the formation of many electron pairs and 
their condensation  
is not at all trivial, 
and it is desirable to clarify this point   
without relying on  
approximation methods. 

One of the results in this direction was
obtained by Shen and Qiu~\cite{SQ93}, 
who proved 
the existence of off-diagonal long-range order, 
closely related to a condensation, 
in the Hubbard model with on-site attractive interactions
on certain bipartite lattices. 
Another was obtained 
by Essler, Korepin, and Schoutens~\cite{Essler93}, 
who proposed
solvable extended Hubbard models exhibiting
a condensation of electrons forming the $\eta$ pair~\cite{comment3,Yang89} 
which is a superposition
of on-site pairs with momentum 0.
There are some rigorous results 
related to these models, in particular,
in one dimension where the models are exactly solvable 
by Bethe ansatz
(see, for example,  \cite{BKS95,KO02,GG03} and references therein). 
Results in higher dimensions are limited. 

In this letter we introduce a new class of 
exactly solvable models exhibiting
a condensation at zero temperature~\cite{Yamanaka03,Tanaka03}.
The models are defined in arbitrary dimensions
and consist of hopping terms of electrons and
local attractive interactions between electrons. 
We show that the models have
the ground states in which all electrons form pairs of
a certain type.
It is worth to note that our pair, unlike the $\eta$ pair
used in the results mentioned above, 
is made up of spin-singlet pairs of electrons at different sites
in addition to on-site pairs.
We also show that, in the momentum space,
the exact ground states are represented 
as superpositions of products of pairs formed {from} electrons
with opposite wave vectors and spins.

As seen later, 
the present models are related to nearly-flat-band models,
which are proved 
to exhibit ferromagnetism~\cite{Mielke,Tasaki95,Tasaki03,TU03}.
Several authors~\cite{Tamura02,Arita02,Kusakabe03} proposed possibilities 
of experimental realization
of the nearly-flat-band ferromagnetism.
Our results suggest that 
there is a chance of finding superconductivity
or interesting phenomena caused by competition
between magnetism and superconductivity as well as ferromagnetism
in the proposed systems.

To simplify the discussion, 
we describe the results in the simplest version of the models.
We can consider similar models on other lattices 
constructed by using the same methods as 
that by Mielke~\cite{Mielke} and that by Tasaki~\cite{Tasaki03}, 
which were developed in the study of (nearly) flat-band ferromagnetism.  
Let $V$ be a $d$-dimensional hypercubic lattice $[1,L]^d\cap\Bbb{Z}^d$
with periodic boundary conditions, 
where $L$ is an arbitrary integer.
Let $M$ be a collection of sites located at the mid-points
of nearest-neighbor pairs in $V$.
Then we define
decorated lattice $\La=V\cup M$.
We denote by $\cis$ and $\cisd$ the annihilation and the creation
operators, respectively, for an electron with spin $\sigma=\up,\dn$ 
at site $i\in\La$. 
These operators satisfy usual fermion anticommutation relations.
The number operator $\nis$ 
is defined as $\nis=\cisd\cis$.
We denote by $\Ne$ the electron number and by $\Phi_0$ 
a state without electrons.
We define fermion operators, which play an essential role in our model, as 
\begin{equation}
 \axs=c_{x,\sigma}-\alpha\sum_{u\in M;|u-x|=1/2}c_{u,\sigma}
\end{equation}
for $x\in V$ and
\begin{equation}
 \bus=c_{u,\sigma}+\alpha\sum_{x\in V;|x-u|=1/2}c_{x,\sigma}
\end{equation}
for $u\in M$, where $\alpha$ is a real number. 
We note that localized single-electron states corresponding
to $a$-operator and $b$-operator are 
orthogonal because 
\begin{equation}
\label{eq:anticommutation1}
\{b_{u,\sigma},a_{x,\sigma}^\dagger\}=0
\end{equation}
for any $x\in V$ and any $u\in M$.

We consider the tight-binding model of electrons
with local attractive interactions 
described by the 
Hamiltonian $H=\Hahop + \Hbhop + \Hint$ with
\begin{eqnarray}
 &&\Hahop=s\sum_{x\in V}\sum_{\sigma=\up,\dn} \axsd\axs,\\
 && \Hbhop=t\sum_{u\in M}\sum_{\sigma=\up,\dn} \busd\bus,\\
 &&\Hint=-W\sum_{x\in V}a_{x,\up}^\dagger a_{x,\dn}^\dagger   
  a_{x,\dn} a_{x,\up},
\end{eqnarray}
where $s,t$ and $W$ are positive parameters.
It is noted that $H$ conserves the electron number $\Ne$
and possesses the spin SU(2) symmetry. 
\begin{figure}
 \includegraphics[width=.3\textwidth]{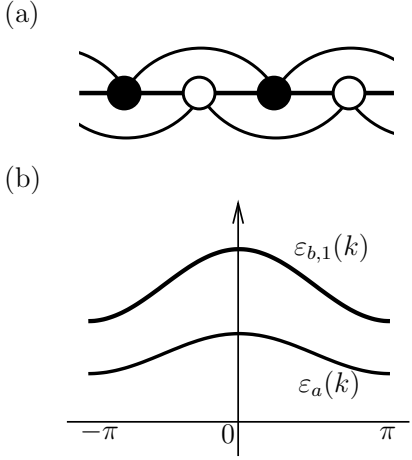}
\caption{The model in one dimension. (a) The lattice structure and
hopping matrix elements. The solid and open circles represent
sites in $V$ and $M$, respectively.
The on-site potential $t_{ii}$ is $s+2\alpha^2t$ 
if $i\in V$ and $2\alpha^2s+t$ if $i\in M$.
The hopping amplitudes are given by 
$t_{ij}=\alpha^2t$ 
if $|i-j|=1,i,j\in V$, 
$t_{ij}=\alpha^2s$ if $|i-j|=1,i,j\in M$,   
$t_{ij}=-\alpha(s-t)$ if $|i-j|=1/2$, and $t_{ij}=0$ otherwise.
(b) The dispersion relations.
The operator $a_{x,\sigma}^\dagger$  
creates an electron in the band with
the dispersion relation 
$\varepsilon_a(k)=s+2\alpha^2s(1+\cos k)$
and $b_{u,\sigma}^\dagger$ creates an electron in the band
with $\varepsilon_{b,1}(k)=t+2\alpha^2t(1+\cos k)$.
}
\end{figure}

The sum of $\Hahop$ and $\Hbhop$ can be reduced to
a standard tight-binding Hamiltonian 
$\sum_{i,j\in\La}\sum_{\sigma}t_{ij}c_{i,\sigma}^\dagger c_{j,\sigma}$,
describing quantum mechanical motion of electrons in $\La$. 
It follows {from} \eqref{eq:anticommutation1} that 
the dispersion relations of the model
are classified into two types; one corresponds to single-electron states
$(\axsd\Phi_0)_{x\in V}$,
and the others correspond to $(\busd\Phi_0)_{u\in M}$. 
More precisely, the dispersion relations are given by
$\varepsilon_{a}(k)=s+2\alpha^2s\sum_{l=1}^d(1+\cos k_l)$,
$\varepsilon_{b,1}(k)=t+2\alpha^2t\sum_{l=1}^d(1+\cos k_l)$,
and $\varepsilon_{b,m}(k)=t$ with $m=2,\dots,d$
(see Fig. 1 for an example in one dimension).
Here, $k$ is the wave vector in the set~\cite{comment4} 
\begin{equation}
 \calK=\left\{
	(2\pi n_1/L ,\dots,2\pi n_d/L)|n_l=0,\pm1,\dots,\pm(L-1)/2 
	\right\}.  
\end{equation}
The energy eigenstate with eigenvalue $\varepsilon_{a} (k)$
is given by
$a_{k,\sigma}^\dagger \Phi_0$ 
with~\cite{symbol}
\begin{equation}
 a_{k,\sigma}^\dagger=\sqrt{\frac{s}{\varepsilon_a(k)|V|}}
  \sum_{x\in V}\rme^{\rmi k\cdot x}\axsd.
\end{equation}
The many-electron ground state of $\Hahop+\Hbhop$ 
is usually the Fermi sea~\cite{comment2}.

By rewriting each term in $\Hint$ as
\begin{equation}
-W\norm{a}^4+W\norm{a}^2a_{x,\up}a_{x,\up}^\dagger
+W a_{x,\dn}a_{x,\up}^\dagger a_{x,\up} a_{x,\dn}^\dagger
\end{equation}
with $\norm{a}^2=(1+2d\alpha^2)$,
one finds that this is bounded below by $-W\norm{a}^4$, 
which is attained by states in the form of 
$a_{x,\up}^\dagger a_{x,\dn}^\dagger\cdots\Phi_0$.   
The terms in $\Hint$ are thus interpreted 
as attractive interactions 
between electrons in localized single-electron states 
corresponding to $a$-operator.
In the $k$-space, $\Hint$ is represented as
\begin{eqnarray}
 \Hint = -\frac{1}{|V|}\sum_{k,k^\prime,q\in \calK}W_{k,k^\prime,q}
 a_{k+q,\up}^\dagger a_{k^\prime-q,\dn}^\dagger a_{k^\prime,\dn} a_{k,\up},\\
 W_{k,k^\prime,q}=\frac{W}{s^2}
  {\sqrt{\varepsilon_a(k+q){\varepsilon}_a(k^\prime-q)
         \varepsilon_a(k^\prime){\varepsilon}_a(k)}},
\end{eqnarray}
which has a similar form to 
the BCS interaction Hamiltonian~\cite{comment5},
although there is a difference in the dependence of
the amplitudes on the wave vectors.
It is noted that interaction $\Hint$ contains
scattering processes 
of electron pairs with non-zero total momentum,
which are usually dropped in the BCS theory.
Nevertheless, our theorem below establishes 
that the ground state of $H$ with fixed even $\Ne$
consists of electron pairs having total momentum~0
(see~\eqref{eq:ground-state} and ~\eqref{eq:zeta4}). 

Before stating our main result, we need to introduce further notation.
Let $G$ be the $|V|\times|V|$ Gram matrix 
for $a$-operator,
whose matrix elements are defined
by $(G)_{xy}=\{a_{x,\sigma},a_{y,\sigma}^\dagger\}$ for $x,y\in V$.
Since $(\axsd\Phi_0)_{x\in V}$ is linearly independent, $G$ is regular
and has its inverse matrix $G^{-1}$.
Then we introduce the dual operators of $a$-operator as
$
 \taxs=\sum_{y\in V}(G^{-1})_{xy}a_{y,\sigma}.
$
It is easy to check the inverse relation
$
 a_{x,\sigma} = \sum_{y\in V}(G)_{xy}\tilde{a}_{y,\sigma}
$
and the anticommutation relations
\begin{equation}
\label{eq:anticommutation2}
 \{\taxs,a_{y,\tau}^\dagger\}=
 \{\axs,\tilde{a}_{y,\tau}^\dagger\}=\delta_{xy}\delta_{\sigma\tau}
\end{equation}
for $x,y\in V$ and $\sigma,\tau=\up,\dn$.
Let us define
\begin{equation}
 \zeta^\dagger = \sum_{x,y\in V}(G)_{xy}
  \tilde{a}_{x,\up}^\dagger \tilde{a}_{y,\dn}^\dagger, 
\end{equation}
which creates two electrons in a spin-singlet pairing state.
We note that $\zeta^\dagger$ is rewritten as 
\begin{equation}
\label{eq:zeta2}
\zeta^\dagger=
\sum_{x\in V}\tilde{a}_{x,\up}^\dagger a_{x,\dn}^\dagger
=\sum_{x\in V}{a}_{x,\up}^\dagger \tilde{a}_{x,\dn}^\dagger  
=\sum_{x,y\in V}
(G^{-1})_{xy}{a}_{x,\up}^\dagger {a}_{y,\dn}^\dagger
\end{equation}
since $G$ and $G^{-1}$ are real symmetric matrices. 
Our main result in this letter is summarized as follows:\\
\textit{Theorem. Consider Hamiltonian $H$ with $W=2s/\norm{a}^2$ and
fixed electron number $\Ne$ in $\Ne\le 2|V|$.
For even $\Ne$, the ground state $\PhiGNe$ is unique, has zero energy, and
is expressed as 
\begin{equation}
\label{eq:ground-state}
 \PhiGNe = \left(\zeta^\dagger\right)^\frac{\Ne}{2}\Phi_0.
\end{equation} 
When $\Ne$ is odd, the ground state has positive energy.}

It is noted that the ground state energy for odd $\Ne$ may 
converge to zero as $L\to\infty$.
Whether this is the case or not should be clarified in a future
study.

If we set $s\le0$ and $W=0$, Hamiltonian $H=\Hahop+\Hbhop$ is
equal to the hopping term of
the nearly-flat-band model
studied in Ref.~\cite{Tasaki95}.
Thus, in this case, the model has saturated ferromagnetic ground states
for $\Ne=|V|$ 
when sufficiently large on-site repulsion is added.  
We note that the ferromagnetic ground states 
found in the above situation
survive for all $W<0$ and also
for sufficiently small positive values of $W$~\cite{comment1}.   
The present model with the on-site repulsion
is expected to describe a quantum phase transition
between magnetic states and electron-pairing states.

The elements of Gram matrix $G$ 
are explicitly given by 
$(G)_{xy}=1+2d\alpha^2$ if $x=y$,
$(G)_{xy}=\alpha^2$ if $|x-y|=1$,
and 
$(G)_{xy}=0$ otherwise, and 
a standard Fourier analysis yields 
$(G^{-1})_{xy}=|V|^{-1}
\sum_{k\in\calK}\frac{s}{\varepsilon_a(k)}
\rme^{\rmi k\cdot(x-y)}$.
{From} this, we find that the pair creation operator $\zeta^\dagger$ 
is written as 
\begin{equation}
\label{eq:zeta4}
 \zeta^\dagger = \sum_{k\in\calK}
  a_{k,\up}^\dagger a_{-k,\dn}^\dagger.
\end{equation}
Substituting this expression into~\eqref{eq:ground-state}, 
one finds that the ground state 
is a superposition of products of 
pairs $a_{k,\up}^\dagger a_{-k,\dn}^\dagger$.  
To see the representation of $\zeta^\dagger$
in terms of $c$-operator, 
we substitute $\axsd=\sum_{i\in\La}\varphi_{xi}\cisd$
into the final expression in \eqref{eq:zeta2},
where $\varphi_{xi}=\{\axs,\cisd\}$.
Then we obtain
\begin{equation}
\label{eq:zeta3}
\zeta^\dagger 
 = \sum_{i\in\La}w_{ii}c_{i,\up}^\dagger c_{i,\dn}^\dagger
 +\sum_{i,j\in\La;i>j}w_{ij}
 (c_{i,\up}^\dagger c_{j,\dn}^\dagger + c_{j,\up}^\dagger c_{i,\dn}^\dagger)
\end{equation} 
with $w_{ij}=\sum_{x,y\in V}(G^{-1})_{xy}\varphi_{xi}\varphi_{yj}$.
Each term in the second sum in~\eqref{eq:zeta3} 
corresponds to a spin-singlet pair of electrons at different sites.  
In the case of $x,y\in V$, 
the weight $w_{xy}$ of a pair reduces to $(G^{-1})_{xy}$,
which has the Ornstein-Zernicke behavior for $|x-y|\gg1$. 

By making a linear combination of
$\PhiGNe$ with different electron numbers, 
we can construct an explicit 
electron-number symmetry breaking ground state.
To see this, let us introduce order parameter 
$\order=\zeta/|V|$.
A straightforward calculation yields
\begin{equation}
\bkt{\order\orderd}_{\La,\Ne}
=\frac{(\PhiGNe,\order \orderd \PhiGNe)}
          {(\PhiGNe,\PhiGNe)}
=\mu_{\La,\Ne}
\end{equation}
with $\mu_{\La,\Ne}=(\Ne/(2|V|)+1/|V|)(1-\Ne/(2|V))$.
Taking a limit $L,\Ne\to\infty$ so that 
the electron density $\Ne/(2|\La|)$ will converge to $\nu$,
we conclude that
$\bkt{\order\orderd}_{\La,\Ne}\to\mu_\nu$
with $\mu_\nu=(d+1)\nu(1-(d+1)\nu)$.
Then, we define 
$
 \PhiGNe^\prime =\left(1+
         {\orderd}/
         {\sqrt{\bkt{\order\orderd}_{\La,\Ne}}}\right)\PhiGNe, 
$
which is also a zero-energy state of $H$.
It is easy to see that
\begin{equation}
\bkt{\order}_{\La,\Ne}^\prime
=\frac{(\PhiGNe^\prime,\order\PhiGNe^\prime)}
          {(\PhiGNe^\prime,\PhiGNe^\prime)}
\to \sqrt{\mu_{\nu}}/2, 
\end{equation}
which implies the explicit electron-number symmetry breaking
for $0<\nu<1/(d+1)$.  

\textit{Proof of the theorem.}
By using $W=2s/\norm{a}^2$, we rewrite $H$
as $H=\Hbhop+\Hintp$ with
\begin{equation}
 \Hintp
  =\frac{W}{2}\sum_{x\in V}\sum_{\sigma=\up,\dn}
  a_{x,\sigma}^\dagger a_{x,-\sigma} 
  a_{x,-\sigma}^\dagger a_{x,\sigma}.
\end{equation}
(Here $-\sigma$ denotes spin opposite to $\sigma$.)
This expression is crucial in our proof.
Since both $\Hbhop$ and $\Hintp$ are 
positive semidefinite operators, one can conclude that 
a zero-energy state of both of these terms is a ground state.
We first show that $\PhiGNe$ in~\eqref{eq:ground-state} is 
in fact a zero-energy state of $H$.
{From} \eqref{eq:anticommutation2}
and $(\axsd)^2=0$, one finds that 
\begin{equation}
 a_{x,\dn}^\dagger a_{x,\up}\zeta^\dagger
  =\zeta^\dagger a_{x,\dn}^\dagger a_{x,\up} 
  +a_{x,\dn}^\dagger a_{x,\dn}^\dagger
  =\zeta^\dagger a_{x,\dn}^\dagger a_{x,\up}.  
\end{equation}
Similarly, we have 
$a_{x,\up}^\dagger a_{x,\dn}\zeta^\dagger
= \zeta^\dagger a_{x,\up}^\dagger a_{x,\dn}$.
These two relations imply that $\Hintp\PhiGNe=0$.
Furthermore it immediately follows {from} \eqref{eq:anticommutation1}
that $\Hbhop\PhiGNe=0$.
Therefore $\PhiGNe$ is a ground state having zero energy.

To prove the other statement 
in the theorem, we use the following lemma, which will be proved
later. \\
\textit{Lemma.  Any zero-energy state $\Phi$ of $H$ with $\Ne\le 2|V|$
(where $\Ne$ is not fixed)
is written as
\begin{equation}
 \Phi=\sum_{A\subset V}\phi_A
  \left(\prod_{x\in A} a_{x,\up}^\dagger \right)
  \left(\prod_{x\in A} \tilde{a}_{x,\dn}^\dagger \right)\Phi_0,
\end{equation}
where coefficients $\phi_A$ satisfy $\phi_A=\phi_{A^\prime}$
if $|A|=|A^\prime|$.}\\
This lemma implies that the ground state energy for fixed odd $\Ne$
is positive.
Let us prove the uniqueness of the ground state
for fixed even $\Ne$.
Now suppose that there are two zero-energy states.
Then any linear combination of these states is also a zero-energy state,
which must satisfy the statement in the lemma.
However, this implies that all the coefficients are vanishing,   
since we can make a suitable linear combination such that
a coefficient $\phi_{A_0}$ for a subset $A_0$ is zero.
This is contradicting with the assumption, and 
thus the ground state, which must have zero energy, is unique.
This completes the proof of the theorem. \qed

\textit{Proof of the lemma.}
Let $\Phi$ be an arbitrary zero-energy state
of $H$ with $\Ne\le 2|V|$.
Since each term $t\busd\bus$ in $\Hbhop$ is positive semidefinite,
$\Phi$ should satisfy $\bus\Phi=0$ for any $u\in M$ and
$\sigma=\up,\dn$.
Thus $\Phi$ must be of the form 
\begin{equation}
 \Phi=\sum_{A_\up,A_\dn\subset V}
  \phi_{(A_\up,A_\dn)}
 \left(\prod_{x\in A_\up} a_{x,\up}^\dagger \right)
 \left(\prod_{x\in A_\dn} a_{x,\dn}^\dagger \right)\Phi_0,
\end{equation}
where $\phi_{(A_\up,A_\dn)}$ are suitable coefficients. 

To be a zero-energy state, $\Phi$ must furthermore 
satisfy $\Hintp\Phi=0$,
i.e.,
$a_{x,-\sigma}^\dagger a_{x,\sigma}\Phi=0$ for any $x\in V$
and $\sigma={\up,\dn}$.
By using 
$
 {\tilde{\Phi}_{(A_\up,A_\dn)}^\up
  =\left(\prod_{x\in A_\up} \tilde{a}_{x,\up}^\dagger \right)
  \left(\prod_{x\in A_\dn} a_{x,\dn}^\dagger \right)\Phi_0},
$
we expand $\Phi$ as
\begin{equation}
\label{eq:expression1}
  \Phi=\sum_{A_\up,A_\dn\subset V} \tilde{\phi}_{(A_\up,A_\dn)}^\up
   \tilde{\Phi}_{(A_\up,A_\dn)}^\up,
\end{equation} 
where $\tilde{\phi}_{(A_\up,A_\dn)}^\up$ are new coefficients,
and operate $a_{x,\dn}^\dagger a_{x,\up}$ on $\Phi$
in this form. 
Then, by using \eqref{eq:anticommutation2}, we have
\begin{eqnarray}
 \sum_{A_\up,A_\dn\subset V}
 \chi[x\in A_\up,x\notin A_\dn]
 \sgn[x;A_\up,A_{\dn}]
 \nonumber\\
 \times\tilde{\phi}_{(A_\up,A_\dn)}^\up
  \tilde{\Phi}_{(A_\up\bs \{x\},A_\dn\cup \{x\})}^\up
  =0,
\end{eqnarray}
where $\sgn[\cdots]$ is a sign factor arising {from} exchanges
of fermion operators, and $\chi[\textrm{``event''}]$ takes 1 if
``event'' is true and takes 0 otherwise.
Since all the terms in the left hand side in the above equation
are linearly independent,
we obtain $\tilde{\phi}_{(A_\up,A_\dn)}^\up=0$ 
if both $x\in A_\up$ and $x\notin A_\dn$.
Since this must hold for all $x\in V$, we find that the sum in 
\eqref{eq:expression1} is restricted to subsets $A_\up, A_\dn$ 
such that $A_\up\subset A_\dn$.  

By using 
$
 \tilde{\Phi}_{(A_\up,A_\dn)}^\dn
  =\left(\prod_{x\in A_\up} {a}_{x,\up}^\dagger \right)
  \left(\prod_{x\in A_\dn} \tilde{a}_{x,\dn}^\dagger \right)\Phi_0,
$
and by taking account of the above result,
we again rewrite $\Phi$ as
\begin{equation}
 \label{eq:expression2}
  \Phi=\sum_{A_\up,A_\dn\subset V;|A_\up|\le|A_\dn|} 
  \tilde{\phi}_{(A_\up,A_\dn)}^\dn
   \tilde{\Phi}_{(A_\up,A_\dn)}^\dn
\end{equation} 
with new coefficients $\tilde{\phi}_{(A_\up,A_\dn)}^\dn$.
Operating $a_{x,\up}^\dagger a_{x,\dn}$ on
$\Phi$ in expression~\eqref{eq:expression2} and repeating
a similar argument to the above,
we find that $\tilde{\phi}_{(A_\up,A_\dn)}^\dn$ must be zero
if $A_\up\ne A_\dn$.

Thus we have shown that $\Phi$ must
be written as 
\begin{equation}
\label{eq:expression3}
 \Phi = \sum_{A\subset V}\phi_A \tilde{\Phi}_{(A,A)}^\dn,
\end{equation}
where $\phi_A = \tilde{\phi}_{(A,A)}^\dn$.    
Now we again operate 
$a_{x,\dn}^\dagger a_{x,\up}
=\sum_{y,y^\prime\in V}(G)_{xy}(G)_{xy^\prime}
\tilde{a}_{y,\dn}^\dagger \tilde{a}_{y^\prime,\up}$ on
$\Phi$ in expression~\eqref{eq:expression3} 
and 
derive conditions on $\phi_A$.
The resulting equation is
\begin{eqnarray}
\sum_{A\subset V}\sum_{y,y^\prime\in V}
\chi[y\notin A,y^\prime\in A]\sgn[y,y^\prime;A]\nonumber\\
\times F_{yy^\prime}^x \phi_A
\tilde{\Phi}_{(A\bs \{y^\prime\},A\cup\{y\})}^{\dn}=0,
\end{eqnarray}
where $F_{yy^\prime}^x = (G)_{xy}(G)_{xy^\prime}$ 
and $\sgn[\cdots]$ is a fermion-sign factor.
Let us choose a subset $A$ which contains site $x$ and does not
contain a site $y$ with $|y-x|=1$. 
By checking the coefficient of 
$\tilde{\Phi}_{(A\bs \{x\},A\cup\{y\})}^\dn$,
we have
\begin{equation}
F_{yx}^x(\sgn[y,x;A]\phi_A+\sgn[x,y;A_{x\to y}]\phi_{A_{x\to y}})=0,
\end{equation}
where $A_{x\to y} = (A\bs\{x\})\cup \{y\}$.
Since $F_{yx}^x$ for ${|x-y|=1}$ is non-zero by definition 
and $\sgn[y,x;A]=-\sgn[x,y;A_{x\to y}]$, we find that
the coefficients must satisfy
$\phi_A=\phi_{A_{x\to y}}$. 
Repeating the same argument for all $x\in V$,
we conclude 
that $\phi_A=\phi_{A^\prime}$ whenever 
$|A|=|A^\prime|$.
This completes the proof of the lemma. \qed

I would like to thank Masanori Yamanaka for useful discussions.


\begin{thebibliography}{99}
\bibitem{comment3}
	The $\eta$-pairing mechanism is originally 
	proposed by Yang in Ref.~\cite{Yang89}, where
	the $\eta$ pair is defined to be a superposition
	of on-site pairs with momentum $\pi$.
\bibitem{comment4}
	We assume that $L$ is an odd integer here.
\bibitem{comment2}
	The dispersion relations for $d\ge2$ 
	contain a flat band,
	and thus the ground states in the non-interacting case are highly
	degenerate when the Fermi energy equals the flat-band energy.
\bibitem{comment1}
	When $s<0$,    
        the model with $\Ne=|V|$ is proved to 
	exhibit saturated ferromagnetism
	for sufficiently large negative values of $W$
	even in the absence of on-site repulsion.
\end{thebibliography}
\end{document}